\documentclass[conference]{IEEEtran}
\IEEEoverridecommandlockouts

\usepackage{paradise}
\usepackage{booktabs}
\usepackage{graphicx}
\usepackage{natbib}
\newcommand{\hypertarget}[2]{}
\newcommand{\texorpdfstring}[2]{#1}
\setcitestyle{numbers,square,comma}

\def\longtable{\begin{table}\let\oldcap\caption%
    \renewcommand\caption[1]{\global\def\captext{##1}}\centering\tabular}
\def\endlongtable{\endtabular \let\caption\oldcap \vskip.5ex \caption{\captext} \end{table}}
\let\endfirsthead\iffalse
\let\endhead\fi

\begin{document}
\bibliographystyle{plainnat}

\title{Using Off-the-Shelf Exception Support Components in C++ Verification%
  \thanks{This work has been partially supported by the Czech Science Foundation grant No. 15-08772S and by Red Hat, Inc.}}

\author{%
\IEEEauthorblockN{Vladimír Štill}
\IEEEauthorblockA{\fimuninarrow \\ Email: xstill@fi.muni.cz}
 \and \IEEEauthorblockN{Petr Ročkai}
\IEEEauthorblockA{\fimuninarrow \\ Email: xrockai@fi.muni.cz}
 \and \IEEEauthorblockN{Jiří Barnat}
\IEEEauthorblockA{\fimuninarrow \\ Email: barnat@fi.muni.cz}
}

\maketitle

\begin{abstract}
  An important step toward adoption of formal methods in software development is support for mainstream programming languages. Unfortunately, these languages are often rather complex and come with
  substantial standard libraries. However, by choosing a suitable intermediate language, most of the complexity can be delegated to existing execution-oriented (as opposed to verification-oriented)
  compiler frontends and standard library implementations. In this paper, we describe how support for C++ exceptions can take advantage of the same principle. Our work is based on \divm{}, an \llvm{}-derived,
  verification-friendly intermediate language.
  
  Our implementation consists of 2 parts: an implementation of the \texttt{libunwind} platform API which is linked to the program under test and consists of 9 C functions. The other part is a
  preprocessor for \llvm{} bitcode which prepares exception-related metadata and replaces associated special-purpose \llvm{} instructions.
\end{abstract}


\section{Introduction}\label{sec:introduction}

Today, formal verification methods are not commonly used in software development, even though they are superior to traditional testing approaches in many respects. One particular example is model
checking, which can be used to control non-determinism in programs, especially when it arises from parallelism. Formal methods can also be used to extend testing coverage (e.g.~via systematic fault
injection), verification of liveness properties or verification of global safety properties (such as global assertions). However, to make those advantages actually available to software developers,
verification tools must be easy to integrate into existing workflows. If the use of verification tools requires substantial effort (as compared to testing), the costs associated with formal methods
can outweigh the savings they provide. This is especially true for modern development processes (especially in commodity software), where there is little time for a separate modeling and design
phases.

For this reason, both the academic and industrial communities~\citep{beyer16:reliab.reprod} increasingly seek to develop and use tools which work with mainstream programming languages. However,
support for these programming languages -- especially when compared to special-purpose modelling formalisms -- brings new complexity to verification tools. Programs written in such languages are
usually more complex and on a lower level of abstraction than models specifically built for analysis tools. Additionally, many programming languages contain features with no counterparts in a typical
modeling language, such as dynamic memory, run-time type information and introspection (RTTI), exception handling, or template instantiation. Moreover, programs written in these languages usually make
use of extensive standard libraries. Therefore, the verifier either has to include all of the language and library functionality as primitives, or it has to provide an implementation which is added to
the verified program just like a traditional library.

The paper is structured as follows: the remainder of Section~\ref{sec:introduction} gives motivation, context and contribution of this work. Section~\ref{sec:exceptions} describes the mechanisms that
C++ implementations typically use in order to support exceptions. The following Section~\ref{sec:execution} then details how \llvm{} is interpreted in \divine{} 4, in particular the parts relevant to
exception handling, such as the stack layout. Section~\ref{sec:transform} and Section~\ref{sec:unwinder} discuss the new components: the \llvm{} transformation and the unwinder, respectively.
Section~\ref{sec:related} surveys the related work and finally, in Section~\ref{sec:evaluation}, we evaluate our approach and we summarise our findings in Section~\ref{sec:conclusion}.

\subsection{Motivation}\label{sec:motivation}

In many cases, it is impractical to re-implement the entire programming language and its support libraries. Verification tools can, however, take advantage of existing compilers or libraries to deal
with some of the complexity. For example, verification can be substantially simplified by translating the source code into an intermediate representation (IR) using an existing compiler frontend. If
the compiler in question can emit intermediate representation after it has been optimised, the verification result is independent of the correctness of the (complex and error-prone) optimiser: any
problems introduced by the optimiser will be caught by the verification tool.

In case of C++, a suitable frontend is the clang compiler, which uses \llvm{} as its IR. Since \llvm{} can optimise the IR and produce executable code on many platforms from a single optimised IR file, the
verification effort does not need to be repeated for each target platform separately. Of course, the code generator (which is comparatively simple when compared to the platform-neutral optimiser)
still needs to preserve the semantics of the program -- otherwise, it would invalidate the verification result.

As an alternative to re-using finished, execution-oriented components, one could only support a subset of a programming language (i.e.~exclude the parts that are hard to support in a verification
tool). However, this weakens the case for supporting mainstream programming languages: it prevents developers from verifying production code. This is especially true for standard libraries, as
programming without them requires the programmer to implement everything from scratch. Finally, the standard library is often implemented in the programming language it is part of, and is therefore
another good candidate for sharing code with execution-oriented implementations of the language. Unfortunately, upstream implementations of standard libraries usually make extensive use of advanced
language features. Consequently, in order to re-use existing standard library implementations, more complete language support is required in the verifier.

Exceptions are among the features that are both widely used (including by the standard library) and tricky to implement. Their use is, however, also common outside of the standard library: libraries
like \texttt{boost} and application-level code often take advantage of this capability. This is natural, since exceptions simplify error handling and usually require less boilerplate code than any of
the alternatives. Furthermore, even though many C++ standard library implementations can be built without exception support\footnote{There are cases where not using exceptions makes sense: if the
  end-user code makes no use of them but the standard library is compiled with exception support, the requisite metadata tables only serve to increase the size of the compiled program.}, this change
can significantly affect its behaviour (and as such, validity of the verification result). Finally, error handling paths, including exception propagation, are an important target for analysis by
verification tools, as they are both hard to test by more conventional means and likely to contain errors -- this naturally arises from the fact that their purpose is to handle unlikely side cases
which can be hard to accurately reproduce with testing. A model checker, on the other hand, can take advantage of its built-in support for non-determinism to rigorously explore error paths.\footnote{This
  is a form of fault injection. When using a model checker, it is only necessary to modify the function where the error may arise (e.g.~the \texttt{malloc} function may be modified to return a
  \texttt{NULL} pointer non-deterministically). The model checker will then take care of exploring all possible combinations of succeeding and failing memory allocations in the program.}

\subsection{Component Re-Use}\label{component-re-use}

Unfortunately, off-the-shelf components from execution-oriented language kits do not provide a complete toolbox that would allow verification tool developers to simply concentrate on verification. The
difficulties roughly fall into two categories: first, the components interact with each other and with the system for which they were originally designed and second, it is often not at all obvious
which components are suitable for re-use and which are not. When a component C is re-used, all the interfaces it uses must be provided as well. There are 3 basic ways in which this can be arranged:

\begin{enumerate}
\def\labelenumi{\arabic{enumi}.}
\tightlist
\item
  re-use another component, D, which provides this interface; this is only possible if all interfaces D uses are already available or can be provided
\item
  modify component C to avoid its dependency on the interface in question
\item
  re-implement the interface as a new, possibly tool- or verification-specific component
\end{enumerate}

\subsection{Contribution}\label{contribution}

The main contribution of this paper is twofold: first, we identify the components that are best re-used and those which are best re-implemented and show that this decision crucially depends on the
underlying intermediate language. Second, we provide implementations of the components which cannot be re-used in a form that is easy to integrate into both existing and future verification tools. One
of the components works as an \llvm{} transformation pass, and could be used with any \llvm{}-based tool. The other component targets the \divm{} language~\citep{rockai18:divm} specifically, and will therefore
only work with tools which understand this language.\footnote{\divm{} is a relatively small extension of the \llvm{} IR, therefore extending tools which work with pure \llvm{} to also support \divm{} may be quite
  easy.}

The goal of this paper, especially in the context of our previous work on the topic of C++ exceptions in verification~\citep{rockai16:model.checkin}, is to aid authors of verification tools to
minimise costs and effort associated with inclusion of exception support. Depending on the characteristics of the tool, either the approach described in~\citep{rockai16:model.checkin} or the one in
this paper might be more suitable. Overall, in a verifier which can handle the \divm{} language or equivalent, the approach given in this paper is simpler to implement and more robust. A more detailed
comparison of the two approaches is given in Section~\ref{sec:cmpD3}.

All source code related to this paper, along with more detailed benchmark results and other supplementary material, are available online under a permissive open-source licence.\footnote{\url{https://divine.fi.muni.cz/2017/exceptions}}

\subsection{Implementation}\label{implementation}

Our primary implementation platform is the \divine{} model checker~\citep{barnat13:divine}. The C++ support in \divine{} has several components: first, \divine{} uses clang to translate C++ into \llvm{} IR. As
outlined above, the verifier does not need to handle complex syntactical features of C++ this way. A few verification-specific transformations are done on the \llvm{} IR before it is converted into the
\divm{} language for execution in \divine{}'s Virtual Machine. The VM executes instructions and performs safety checks, such as bound checking. Alone, these components provide basic support for C++. In
order to support features such as RTTI and exceptions, it is also necessary to provide a runtime support library and an implementation of the standard library. These libraries in turn rely on a C
standard library and on a threading library (\texttt{pthreads} on POSIX compatible systems). Those libraries are provided by \dios{}, a small, verification-oriented operating system which runs inside
\divm{}.

As discussed above, building those libraries into the verifier is impractical due to cost and time constraints. There is, however, another important reason why these should be kept out of the
verification core: any extension of the verifier increases risks of implementation errors, and the more complex these extensions are, the higher are the associated risks. Moreover, any such errors in
the verifier can lead to incorrect verification results. For this reason, \divine{} ships source code implementing these libraries as separate modules; this source code is later compiled into \llvm{} IR and
linked to the verified program. This way, the libraries are subject to the same error checking as user code, and any errors in their implementation that are exposed by the user program will be
detected by the verifier.

Additionally, whenever off-the-shelf components are re-used, it is preferable to keep verification-specific changes at minimum. The standard C library in \divine{} is based on PDCLib, a small, portable,
public domain C library. The copy of PDCLib in \divine{} includes a few modifications (the C library interfaces directly with the operating system in many cases, therefore it is necessary to port it to
work with the verifier, much like it would be necessary to port it to a new operating system). For threading support, \divine{} ships with a custom implementation of the \texttt{pthread} library (so far,
no existing implementation of the \texttt{pthread} interface which could be re-used has been identified). For C++ support, \texttt{libc++abi} (the runtime library) and \texttt{libc++} (the standard
library) are used. Both of these libraries are maintained by the \llvm{} project and work on many Unix-like systems.

\begin{figure}
\centering
\includegraphics[width=0.47000\textwidth]{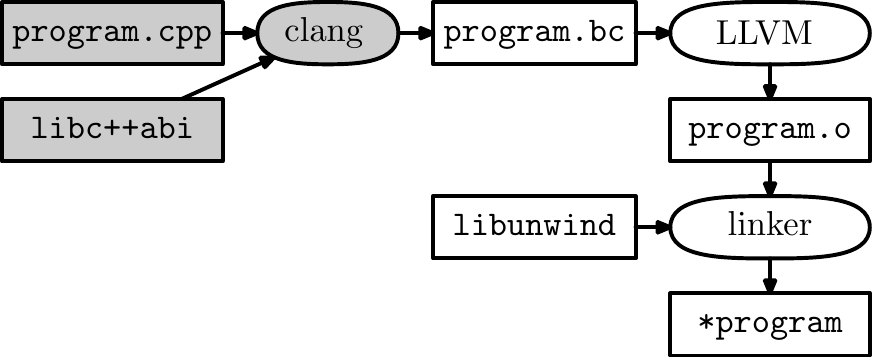}
\caption{Components involved in exception support in the standard clang/\llvm{} stack. Under the scheme proposed in this paper, the highlighted elements are shared between verification and execution
environments.}\label{fig:components}
\end{figure}

\begin{figure}
\centering
\includegraphics[width=0.47000\textwidth]{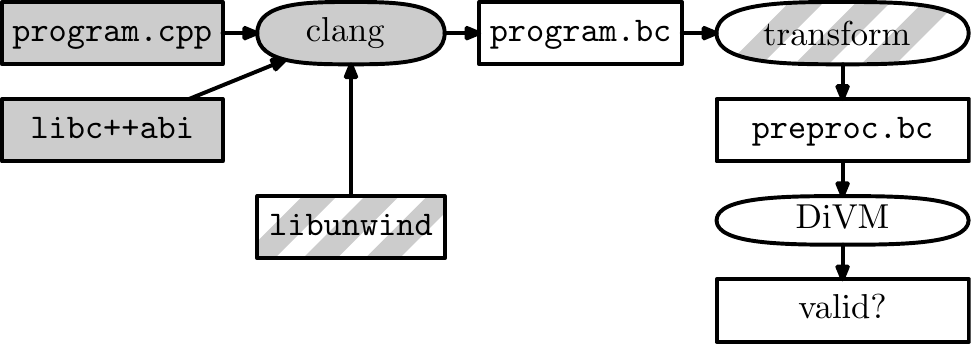}
\caption{Components involved in exception support in the \divine{} 4 C++ verification stack. The solid-filled elements are re-used without modification from the execution-oriented clang/\llvm{} stack (cf.
Figure~\ref{fig:components}). The hatch-filled components are the additions described in this paper.}\label{fig:divine}
\end{figure}

\subsection{Components for Exception Support}\label{components-for-exception-support}

Unlike other features of C++, exceptions are neither handled by the standard or runtime libraries alone, nor delegated to the C standard library (as C has no support for exceptions). Instead,
\texttt{libc++abi} provides exception support with the help of a platform-specific \emph{unwinder library} which is responsible for stack introspection and unwinding (removal of stack frames and
transfer of control to exception-handling code). The interaction of these components is illustrated in Figure~\ref{fig:components}.

For this reason, \divine{} has to either provide an unwinder implementation compatible with \texttt{libc++abi}, or modify \texttt{libc++abi} to use custom code for exception handling. In \divine{} 3, the
latter approach was used, as it was deemed easier at the time~\citep{rockai16:model.checkin}. However, while basic exception support was easier to achieve this way, the approach also had its
disadvantages. First, the \llvm{} interpreter in \divine{} 3 had special support for exception-related functionality. Second, the \texttt{libc++abi} code for exception handling was replaced, which had 2
important consequences: first, the replacement code was not comprehensive enough\footnote{That is, some of the less frequently used features of C++ exceptions were handled either incorrectly or not at
  all. That is to say, the size of the \texttt{libc++abi} portion that would have needed to be re-implemented was initially underestimated.} and second, this meant that the replaced part of
\texttt{libc++abi} was not taken into account during verification.

In this paper, we instead take the first approach: re-use \texttt{libc++abi} in its entirety and provide the interfaces it requires. Therefore, we have implemented the \texttt{libunwind} interface
used by \texttt{libc++abi} for stack unwinding and an \llvm{} transformation which builds metadata tables that \texttt{libc++abi} needs to decide which exceptions should be caught, how they should be
handled and which functions on the stack need to perform cleanup actions. The situation is illustrated in Figure~\ref{fig:divine}.

Using the original \texttt{libc++abi} code means that all features of the C++ exception system are fully supported and verification results also cover the low-level exception support code. That is,
this portion of the code is identical in both the bitcode which is verified and in the natively executing program.\footnote{Clearly, the \texttt{libunwind} implementation is different in those two
  environments, and therefore correctness of the platform-specific implementation of \texttt{libunwind} must be established separately.} Finally, the proposed design is easier to extend to other
programming languages.

\subsection{Other Components in Use}\label{sec:components}

In line with the principles outlined so far, the implementation of the C and C++ standard libraries (and the C++ runtime library) used in \divine{} are third party code with only minimal modifications.
The C standard library implementation (PDCLib) consists of approximately 38 thousand lines of code, while the C++ runtime library (\texttt{libc++abi}) and the C++ standard library (\texttt{libc++})
contain 8 and 12 thousand lines of code, respectively.

Standard libraries inevitably contain platform-specific code, and this is also true of the implementations bundled with \divine{}. The modifications due to the porting effort were, however, quite
minimal, since \dios{} already provides a very POSIX-like interface. The C library was, unsurprisingly, affected the most: changes in memory allocation, program startup- and exit-related functions and in
handling of the \texttt{errno} variable were required. In \texttt{libc++}, however, the changes were limited to platform configuration macros and the only change in \texttt{libc++abi} was a
\dios{}-specific tweak in allocation of thread-local storage for exception handling.

Since user programs and libraries alike rely on the POSIX threading API (also known as \texttt{pthread}), this API is provided by \dios{} and is implemented in about 2000 lines of C++. The
\texttt{libunwind} implementation introduced in this paper brings in additional 350 lines of code (the implementation is done in exception-free C++). Likewise, the C library and everything above also
depends on low-level filesystem access routines provided by the operating system. In \dios{}, this IO and filesystem layer (VFS\footnote{Short for Virtual File System, since in a verification
  environment, the system under test must not access the real filesystem or any other part of the outside environment.}) is implemented in about 5500 lines of C++ code and uses exceptions heavily for
error propagation.

So far, all the components mentioned in this section are linked with the user program to form the final bitcode file for verification. For comparison, the verification core of \divine{} (the \divm{}
evaluator, memory management and the verification algorithm), amounts to roughly 6 thousand lines of C++. Finally, there is about 2500 lines of code which implement various transformations on the \llvm{}
bitcode. Out of these 2500 lines, less than 300 are part of the exception-related extension described in this paper.

\section{Exceptions in C++}\label{sec:exceptions}

Throwing an exception requires removal of all the stack frames\footnote{The execution stack of a (C++) program consists of stack frames, each holding context of a single entry into some function. It
  contains local variables, a return address and register values which need to be restored upon return.} between the throwing and catching function from the stack (\emph{unwinding}). Therefore,
exception handling is closely tied to the particular platform and is described by ABI\footnote{\emph{Application Binary Interface}, a low-level interface between program components on a given
  platform.} for the platform. Commonly, exception handling is split into two parts, one which is tied to the platform (the \emph{unwinder library} which handles stack unwinding) and one which is tied
to the language and provided by the language's runtime library.\footnote{There are many implementations of the C++ runtime library, which, besides exception support code, provides additional features
  such as RTTI. Each implementation is usually tied to a particular C++ standard library. Commonly used implementations on Unix-like systems are \texttt{libsupc++}, which comes with \texttt{libstdc++}
  and the GCC compiler, and \texttt{libc++abi}, which is tied to \texttt{libc++} used by some builds of clang and by \divine{}.} These two parts cooperate in order to provide exception handling for a
given language; however, this communication is not standardised in any cross-platform fashion. For this reason, we will now focus on zero-cost exceptions based on the Itanium ABI, an approach which is
used across various Unix-like systems on \texttt{x86} and \texttt{x86\_64} processor architectures and is the preferred basis for \llvm{} exceptions. Nevertheless, it is possible to generalize our
results to other implementations.

\subsection{Zero-Cost Exceptions}\label{sec:zerocost}

The so-called zero-cost exceptions are designed to incur no overhead during normal execution, at the expense of relatively costly mechanism for throwing exceptions. This in particular means that no
checkpointing is possible. Instead, when an exception is thrown, the exception support library, with the help of \emph{unwind tables}, finds an appropriate \emph{handler} for the exception and uses
the \emph{unwinder} to manipulate the stack so that this handler can be executed. The search for the handler is driven by a \emph{personality function}, which is provided by the implementation of the
particular programming language.

The personality function is responsible for deciding which handler should execute (the handler selection can be complex and language-specific). In general, there are two types of handlers,
\emph{cleanup handlers}, which are used to clean up lexically scoped variables (and call their destructors, as appropriate) and \emph{catch handlers}, which contain dedicated exception-handling code.
The latter typically arise from \texttt{catch} blocks. Another major difference between those two types of handlers is that catch handlers stop the propagation of the exception, while cleanup handlers
let propagation continue after the cleanup is performed. While cleanup handlers are usually run unconditionally, the catch handler to be executed, if any, is determined by the personality
function.\footnote{In fact, the personality function can also decide to skip cleanup handlers, but this is not common.} In C++, the personality function selects the closest \texttt{catch} statement
which matches the thrown type (the match is determined dynamically, using RTTI). The personality function consults the unwind tables, in particular their \emph{language-specific data area (LSDA)}, to
find information about the relevant catch handlers.

When an exception is thrown, the runtime library of the language creates an \emph{exception object} and passes it to the unwinder library. The actual stack unwinding is, on platforms which build on
the Itanium ABI, performed in two phases. First, the stack is inspected (without modification) in search for a catch handler. Each stack frame is examined by the relevant personality
function.\footnote{Different personality functions can be called for different frames, for example if the program consists of code written in different languages with exception support.} If an
appropriate catch handler is found in this phase, unwinding continues with a second phase; otherwise, an unwinder error is reported back to the throwing function. Unwinder errors usually cause program
termination. In the second phase, the stack is examined again, and a personality function is invoked again for each frame. In this phase, cleanup handlers come into play. If any handler is found
(cleanup or catch), this fact is indicated to the unwinder, which performs the actual unwinding to the flagged frame. Once the control is transferred to the handler, it can either perform cleanup and
resume propagation of the exception, or, if it is a catch handler, end the propagation of the exception. If exception propagation is resumed, the unwinder continues performing phase 2 from the point
of the last executed handler. This is facilitated by storing the state of the unwinder within the exception object.

\subsection{Unwind Tables}\label{unwind-tables}

As mentioned in Section~\ref{sec:zerocost}, both the unwinder library and the language runtime depend on unwind tables for their work. The unwinder uses these tables to get information about stack
layout in order to be able to unwind frames from it, and for detection which personality function corresponds to a frame. The personality function then uses the language-specific data area (LSDA) of
these tables in its decision process.

While the unwinder part of the tables is unwinder- and platform-specific (it depends on stack layout), the LSDA is platform- and language-specific. For these reasons, unwind tables are not present in
the \llvm{} IR; instead, they are generated by the appropriate code generator for any given platform, based on information in the \texttt{landingpad} instructions, and the personality attribute of
functions. On Unix-like systems, the unwind tables are in the DWARF\footnote{DWARF is a standard for debugging information designed for use with ELF executables. It is used on most modern Unix-like
  systems.} format.

\section{Execution of \llvm{} programs}\label{sec:execution}

In this section, we will look at how \llvm{} bitcode is executed by a model checker and how this execution is affected by addition of exception support. Unlike previous approaches, the technique
described in this paper does not require any exception-specific intrinsic functions or hypercalls to be supported by the verifier. The exception-specific \llvm{} instructions can be implemented in the
simplest possible way: \texttt{invoke} becomes equivalent to a \texttt{call} instruction followed by an unconditional branch. The \texttt{landingpad} instruction can be simply ignored by the verifier
and \texttt{resume} instructions and calls of the \texttt{llvm.eh.typeid.for} intrinsic are both removed by the transformation described in Section~\ref{sec:transform}. Moreover, the metadata required
by \texttt{libc++abi} are likewise generated by the \llvm{} transformation and this process is completely transparent to the verifier.

In addition to support for \llvm{}, the unwinder (described in more detail in Section~\ref{sec:unwinder}) requires the ability to traverse and manipulate the stack and read and write \llvm{} registers
associated with a given stack frame. Finally, it needs access to a representation of the bitcode for a given function. All those abilities are part of the \divm{} specification~\citep{rockai18:divm} and
are generally useful, regardless of their role in exception support.

The \divm{} implementation in \divine{} 4 handles execution of \llvm{} instructions, \llvm{} intrinsic functions and \divm{}-specific \emph{hypercalls}.\footnote{Intrinsic functions are provided by \llvm{} as a
  light-weight alternative to new instructions. Such functions are recognized and translated by \llvm{} itself, as opposed to ``normal'' functions that come from libraries or the program. Likewise, \divm{}
  provides hypercalls, which are functions that are, in addition to \llvm{} intrinsics, recognized by \divm{}.} Hypercalls exist to allocate memory, perform nondeterministic choice or to set \divm{}'s
\emph{control registers} (which contain, among other, the pointer to the currently executing stack frame). Additionally, \divm{} performs safety checks, such as memory bound checking, and detects use of
uninitialised values. However, \divm{} hypercalls are intentionally low-level and simple and do not provide any high-level functionality, such as threading or standard C library functionality. Instead,
those are provided by the \divine{} Operating System (\dios{}) and the regular C and C++ standard libraries.

The most important purpose of \dios{} is to provide threading support. To this end, \dios{} provides a \emph{scheduler}, which is responsible for keeping track of threads and their stacks and for
(nondeterministically) deciding which thread should execute next. This scheduler is invoked repeatedly by the verifier to construct the state space. The scheduler fully determines the behaviour (or
even presence) of concurrency in the verified program: while \dios{} provides asynchronous, preemptive parallelism typical of modern operating systems, it is also possible to implement cooperative or
synchronous schedulers instead.

\subsection{Stack Layout and Control Registers}\label{sec:stack-layout}

A \divm{} program can have multiple stacks, but only one of them can be active at any given time (a pointer to the active stack is kept in a \divm{} control register). The active stack is normally either
the kernel stack or the stack that belongs to the active thread which was selected by the scheduler. Switching of stacks (and program counters) is performed by the \texttt{control} hypercall which
manipulates \divm{} control registers.

\begin{figure}
\centering
\includegraphics[width=0.47000\textwidth]{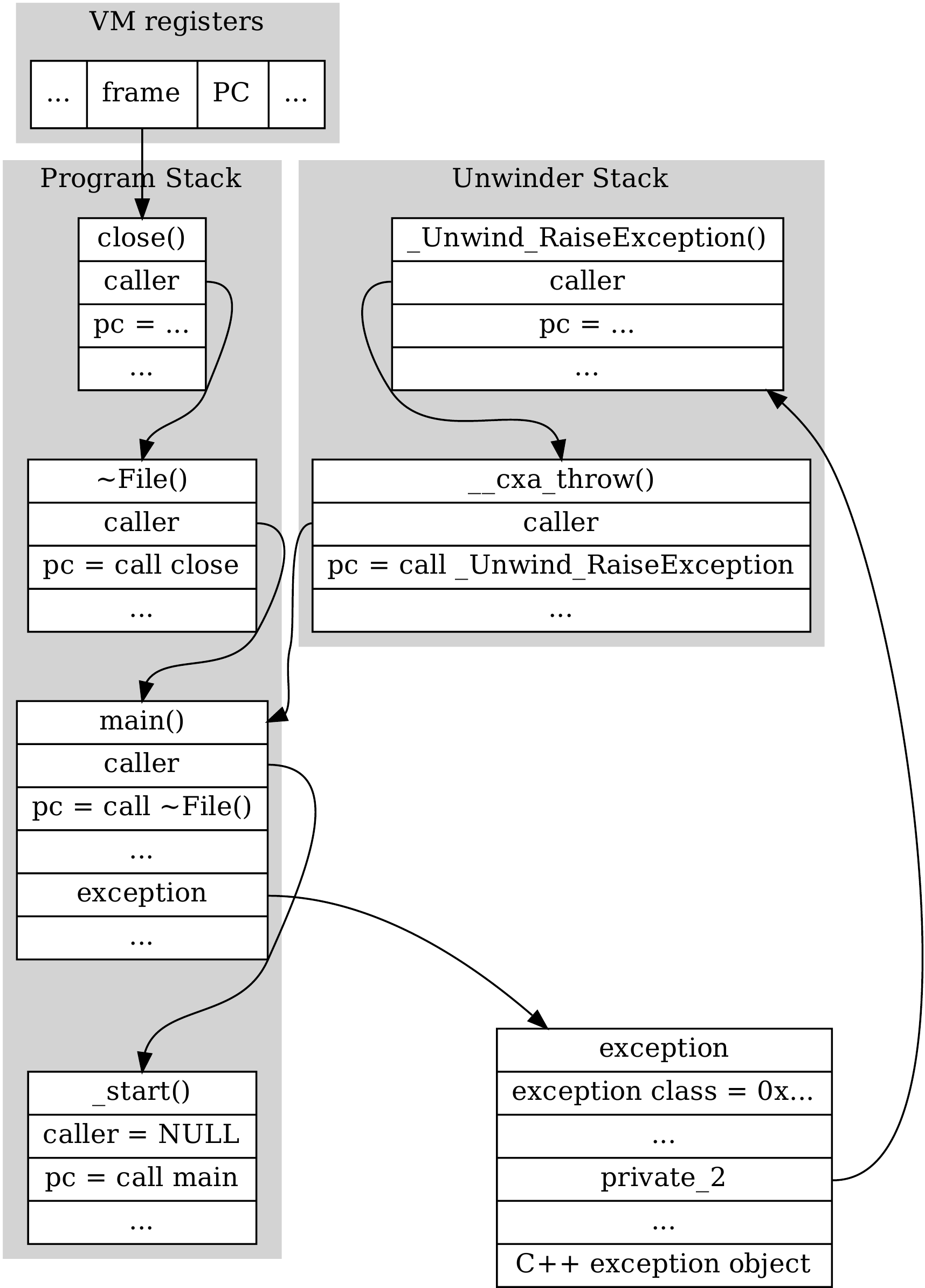}
\caption{In this figure we can see a stack of a program which is running cleanup block in the \texttt{main} function. The cleanup block calls the destructor of \texttt{File} structure, which in turn
calls the \texttt{close} function (which is the current active function). Furthermore, the cleanup handler can access the exception object which contains a pointer to the stack of the unwinder. This
pointer is used by the implementation of the \texttt{resume} instruction to jump back to the unwinder and continue phase 2 of the unwinding.}\label{fig:treestack}
\end{figure}

Traditionally, stack is represented as a continuous block of memory which contains an activation frame for each function call. In \divm{}, the stack is not continuous; instead, it is a singly-linked list
of activation frames, each of which points to its caller. This has multiple advantages: first, it is easy to create a stack frame for a function, for example when \dios{} needs to create a new thread;
additionally, the linked-list-organized stack is a natural match for the graph representation of memory which \divm{} mandates, and therefore can be saved more efficiently~\citep{rockai18:divm}.
Additionally, this way the stack may be nonlinear, and the unwinder can use this feature to safely transfer control to a cleanup block while the unwinder frame is still on the stack. Later, the
handler can return control to the unwinder frame and the unwinder can continue its execution. This would be impossible with a continuous stack since cleanup code is allowed to call arbitrary functions
and frames of those functions would overwrite the frame of the unwinder. For this reason, on traditional platforms, the unwinder needs to store its entire state in the exception object, while in \divm{},
it can simply retain its own activation frame. An illustration of how the stack looks while the unwinder is active is shown in Figure~\ref{fig:treestack}.

\section{The \llvm{} Transformation}\label{sec:transform}

The C++ runtime library (\texttt{libc++abi} in our case), needs access to the LSDA section of unwind tables (a pointer to this metadata section is accessible through the unwinder interface). This
section contains DWARF-encoded exception tables, which are normally generated together with the executable by the compiler backend (code generator). Unfortunately, the generator of DWARF exception
tables in \llvm{} is closely tied to the machine code generator and cannot be used to generate DWARF-formatted exception tables for verification purposes. For this reason, we have implemented a small
\llvm{} transformation which processes the information in \texttt{landingpad} instructions and generates \llvm{} constants which contain the DWARF-formatted LSDA data. A reference to one such constant is
attached to each function in the bitcode file.

To improve efficiency, \llvm{} does not directly use RTTI type info pointers within the landing blocks to decide which exception handlers should run. RTTI objects are special C++ objects which are used
to identify types at runtime and are emitted by the C++ frontend as constants. Due to the complexities of C++ type system, matching RTTI types against each other is expensive: a search in a pair of
directed acyclic graphs is required. Moreover, since the RTTI matching must be already done in the personality function to decide which frames to unwind, the personality can also pre-compute a
numerical index for the landing pad. This index, also called a \emph{selector value} is then used as a shortcut to run an appropriate \texttt{catch} clause within the landing block, instead of
re-doing the expensive RTTI matching. Since the \texttt{catch} handler is typically expressed in terms of typeinfo pointers, it needs to efficiently obtain the selector value from a type info pointer.
For this purpose, \llvm{} provides a \texttt{llvm.eh.typeid.for} intrinsic, which obtains (preferably at compile time) the selector value corresponding to a particular type info pointer.

Therefore, besides generating the LSDA data, the transformation statically computes the values which correspond to \texttt{llvm.eh.typeid.for} calls and substitutes them into the bitcode. Since the
purpose of \texttt{llvm.eh.typeid.for} is to translate from RTTI pointers to selector values, it is only required that the integer selector value chosen for a particular RTTI object is in agreement
with the personality function. In our implementation, this is ensured by computing the selector values statically for both the LSDA (which is where personality function obtains them) and for
\texttt{llvm.eh.typeid.for} at the same time.

Finally, the transformation rewrites all uses of the \texttt{resume} instruction to ordinary calls to \texttt{Resume}, a function which is part of \texttt{libunwind} (see also
Table~\ref{tbl:libunwind}).

\section{The Unwinder}\label{sec:unwinder}

The unwinder in \divine{} is designed around the interface described in the Itanium C++ ABI documentation,\footnote{\url{https://mentorembedded.github.io/cxx-abi/abi.html}} adopted by multiple vendors
and across multiple architectures. The implementation is part of the runtime libraries shipped with \divine{}.\footnote{\texttt{runtime/libc/functions/unwind.cpp}} The unwinder builds upon a lower-level
stack access API which is provided by \dios{} under \texttt{sys/stack.h}.

\hypertarget{tbl:libunwind}{}
\begin{longtable}[]{@{}ll@{}}
\caption{\label{tbl:libunwind}A list of C functions provided by \texttt{libunwind}. In C, all the functions are prefixed with \texttt{\_Unwind\_} to prevent name conflicts with user code and other
libraries (i.e.~the C name of \texttt{SetGR} is \texttt{\_Unwind\_SetGR}). }\tabularnewline
\toprule
Function & Description\tabularnewline
\midrule
\endfirsthead
\toprule
Function & Description\tabularnewline
\midrule
\endhead
\texttt{SetGR} & Store a value into a general-purpose register\tabularnewline
\texttt{GetGR} & Read a value from a general-purpose register\tabularnewline
\texttt{SetIP} & Stora a value into the program counter\tabularnewline
\texttt{GetIP} & Read the value of the program counter\tabularnewline
\texttt{RaiseException} & Unwind the stack\tabularnewline
\texttt{Resume} & Continue unwinding the stack after a cleanup\tabularnewline
\texttt{DeleteException} & Delete an exception object\tabularnewline
\texttt{GetLSDA} & Obtain a pointer to the LSDA\tabularnewline
\texttt{GetRegionStart} & Obtain a base for relative code pointers\tabularnewline
\bottomrule
\end{longtable}

Due to the stack layout used in \divm{} (a linked list of frames, see also Section~\ref{sec:stack-layout}), our unwinder is much simpler than usual. The main task of unwinding is handled by the
\texttt{RaiseException} function, which is called by the language runtime when an exception is thrown. This function performs the two phase handler lookup described in Section~\ref{sec:zerocost} and
it adheres to the Itanium ABI specification, with the following exceptions:

\begin{enumerate}
\def\labelenumi{\roman{enumi}.}
\tightlist
\item
  it checks that an exception is not propagated out of a function which has the nounwind attribute set, and reports verification error if this is the case;
\item
  if the exception is a C++ exception and there is no handler for this exception type, the unwinder chooses nondeterministically whether it should or should not unwind the stack and invoke cleanup
  handlers.
\end{enumerate}

The purpose of the first deviation is to check consistency of exception annotations (arising, for example, from a \texttt{nothrow} function attribute as available in GCC and in clang). The second
modification allows \divine{} to check both allowed behaviours of uncaught exceptions in C++: the C++ standard specifies that it is implementation-defined whether the stack is unwound (and destructors
invoked) when an exception is not caught.\footnote{Section 15.5, paragraph 9 of the C++ standard~\citep{cxx14:standar.program}} Since the program may contain errors which manifest only under one of
these behaviours, it is useful to be able to test both of them.

\subsection{Low-Level Unwinding}\label{sec:unwinder:ll}

The primary function of the unwinder described above is to find exception handlers; for the actual unwinding of frames, it uses a lower-level interface provided by \dios{}. This interface consists of two
functions: \texttt{\_\_dios\_jump}, which performs a non-local jump, possibly affecting both the program counter and the active frame, and \texttt{\_\_dios\_unwind}, which removes stack frames from a
given stack. \texttt{\_\_dios\_unwind} is designed in such a way that it can unwind any stack, not only the one it is running on, and is not limited to the topmost frames (effectively, it removes
frames from the stack's singly-linked list, freeing all the memory allocated for local variables that belong to the unlinked frames, along with the frames themselves\footnote{When a function returns
  normally (due to a \texttt{ret} instruction), \divm{} takes care of freeing the frame and its local variables (\texttt{alloca} memory).}). The unwinder identifies values as local variables by looking
at the instructions of the active function -- the results of \texttt{alloca} instructions are exactly the addresses of local variables.

\subsection{Unwinder Registers}\label{unwinder-registers}

When an exception is propagating, a personality function has to be able to communicate with the code which handles the exception. In C++, the communicated information includes the address of the
exception object and a selector value which is later used by the handler. On most platforms, these values are passed to the handler using registers, which are manipulated using unwinder's
\texttt{SetGR} function. This function can either set the register directly (if it is guaranteed not to be overwritten before the control is transferred to the handler), or save the value in a
platform-specific way and make sure it is restored before the handler is invoked.

In \llvm{} (and hence in \divm{}), there is no suitable counterpart to the general purpose registers of a CPU; instead, the values set by the personality function should be made available to the program in
the return value of the \texttt{landingpad} instruction. This, however, requires the knowledge of the expected semantics of these registers. Currently, all users of the unwinder are expected to use
the same registers as the C++ frontend in clang. That is, register 0 corresponds to the exception object and register 1 corresponds to a type index. This also directly maps to the return type of
\texttt{landingpad} instructions and therefore the register values can be saved directly into the \llvm{} register corresponding to the particular \texttt{landingpad} that is about to be executed.

Registers other than 0 and 1 are currently not supported. In \llvm{}, in line with the above observation about clang and C++, there is a convention that \texttt{SetGR} indices correspond to indices into
the result tuple of a \texttt{landingpad} instruction. As long as this convention is preserved by a particular language frontend and its corresponding runtime library (personality function), it is
very easy to extend our unwinder to support this language. Finally, if a language frontend were instead to emit calls to \texttt{GetGR} in the handler, registers of this type can be stored in the
unwinder \texttt{Context} directly.

\subsection{Atomicity of the Unwinder}\label{atomicity-of-the-unwinder}

The unwinder performs rather complex operations and therefore throwing an exception can create many states, even when \(τ\) reduction~\citep{rockai13:improv.state} is enabled. However, many of these
states are not interesting from the point of view of verification, as the operations performed by the unwinder are mostly thread-local and only the exception handlers (and possibly personality
function) can perform globally visible actions. For this reason, the unwinder uses \divm{}'s atomic sections to hide most of its complexity.

Since an atomic section is implemented as an \emph{interrupt mask} (i.e.~a single flag indicating that an atomic section is executing) in \divm{}, it is necessary to correctly maintain the state of this
flag. In particular, it is required that the unwinder behaves reasonably even if it is called when the program is already in an atomic section. Consequently, care must be taken to restore the state of
the atomic mask when the unwinder transfers control to a personality function or an exception handler. When the unwinder is first called, it enters an atomic section and saves the previous value of
the interrupt mask. This will be the value the flag will be restored to when a personality function is first invoked. The mask is later re-acquired after the personality function returns and it is
restored once more when the first handler is invoked. When the exception handler resumes (using the \texttt{resume} instruction), the atomic section is re-entered and its state saved so its state
before the resume can be restored again for the next call to a personality function. This way, it is possible to safely throw an exception out of an atomic section, provided that the atomic section is
exception-safe (that is, it has an exception handler which ends the atomic section if an exception is propagated out of it).

\subsection{\texorpdfstring{\texttt{longjmp} Support}{longjmp Support}}\label{longjmp-support}

Using the low-level unwinder interface described in Section~\ref{sec:unwinder:ll}, it is easy to implement other mechanisms for non-local transfer of control. The functions \texttt{longjmp} and
\texttt{setjmp}, specified as part of C89, are one such example.\footnote{Implemented in \texttt{runtime/libc/includes/setjmp.h} and \texttt{runtime/libc/functions/setjmp/}.} The \texttt{setjmp}
function can be used to save part of the state of the program, so that a later call to \texttt{longjmp} can restore the stack to the state it was in when \texttt{setjmp} was called. This way,
\texttt{longjmp} can be used to remove multiple frames from the stack. When \texttt{longjmp} is called, the program behaves as if \texttt{setjmp} returned again, only this time it returns a different
value (provided as an argument to \texttt{longjmp}).

The \divine{} implementation of \texttt{setjmp} saves the program counter and the frame pointer of the caller of \texttt{setjmp}. The \texttt{longjmp} function then uses this saved state, along with
access to the text of the program, to set the return value of the \texttt{call} instruction corresponding to the \texttt{setjmp}. Afterwards, it unwinds the stack using the low-level stack access API
from \texttt{sys/stack.h} and transfers control to the instruction right after the call to \texttt{setjmp}.

\section{Related Work}\label{sec:related}

Primarily, we have looked at existing tools which support verification of C++ programs. Existence of an implementation is, to a certain degree, an indication that a given approach is viable in
practice. We have, however, also looked at approaches proposed in the literature which have no implementations (or only a prototype) available.

A number of verification tools are based on \llvm{} and therefore have some support for C++. LLBMC~\citep{sinz12:llbmc} and NBIS~\citep{gunther14:increm.bounded} are \llvm{}-based bounded model checkers
which target mainly C and have no support for exceptions or the C++ standard library. VVT~\citep{gunther16:vienna.verific.tool} is a successor of NBIS which uses either IC3 or bounded model checking
and has limited C++ support, but it does not support exceptions. Furthermore, KLEE~\citep{cadar08:klee} and KLOVER~\citep{li11:klover} are \llvm{}-based tools for test generation and symbolic execution.
KLOVER targets C++ and according to~\citep{li11:klover} has exception support, but it is not publicly available. On the other hand, KLEE focuses primarily on C and its C++ support is rather limited
and it has no exception support.

Both CBMC~\citep{clarke04:tool.checkin, kroening14:cbmc} and ESBMC~\citep{ramalho13:smt.based} bounded model checkers support C++ (but neither appears to support the standard library) and they include
support for exceptions. However, in CBMC, the support for exceptions is limited to throwing and catching fundamental types.\footnote{A simple test which throws and tries to catch an exception object
  crashes CBCM 5.6.} In our survey of tools for verification of C++ programs, ESBMC has by far the best exception support: the latest version can deal with most, but not all\footnote{ESBMC 3.0 is
  unable to determine that an exception ought to be caught when the \texttt{catch} clause specifies a type which is a virtual base class in a diamond-shaped hierarchy and the object thrown is of the
  most-derived type of the diamond. This suggests that ESBMC uses its own implementation of RTTI support code, which is somewhat incomplete, compared to production implementations.}, types of
exception handlers and even with exception specifications. Finally, \divine{} 3~\citep{rockai16:model.checkin} also comes close to full support for exceptions, but lacks support for exception
specifications. Overall, this survey suggests that all current implementations of C++ exceptions in verification tools are incomplete and confirms that using an existing, standards-compliant
implementation in a verification tool is indeed quite desirable.

Finally, it is also possible to transform a C++ program with exceptions into an equivalent program which only uses more traditional control flow constructs. This approach was taken
in~\citep{prabhu11:interp.except}, with the goal of re-using existing analysis tools without exception support. While this approach is applicable to a wide array of verification tools, it is also
incompatible with re-use of existing exception-related runtime library code. As such, it offers a very different set of tradeoffs than our current approach. Moreover, the translation cost is far from
negligible, and also affects code that does not directly deal with exceptions (i.e.~it violates the zero-cost principle of modern exception handling). Unfortunately, we were unable to evaluate this
approach, since there are no publicly available tools which would implement it.

\section{Evaluation}\label{sec:evaluation}

In order to asses the viability of our approach, we have executed a set of benchmarks in various configurations of \divine{}~4. The benchmarks were executed on quad-core Xeon 5130 clocked at 2 GHz and
with 16GB of RAM. We have measured the wall time, making all 4 cores available to the verifier.

\subsection{Benchmark Models}\label{benchmark-models}

The set of models we have used for this comparison consists of 831 model instances, out of which we picked the 794 that do not contain errors. The reason for this is that the execution time is much
more variable when a given program contains an error, since the model checking algorithm works on the fly, stopping as soon as the error is discovered.

Majority of the valid models (777) are C++ programs of varying complexity, while the 17 models in the svc-pthread category are concurrent programs written in plain C with pthreads. Since our
implementation of the pthread API is done in C++, the impact of exception support on verification of C programs is also relevant. The ``alg'' category includes sequential algorithmic and data
structure benchmarks, the ``pv264'' category contains unit tests for student assignments in a C++ course, the ``iv112'' category contains unit tests for concurrent data structures and other parallel
programs (again assignment problems in a C++ course), ``libcxx'' contains a selection of the \texttt{libc++} testsuite (with focus on exception support coverage), ``bricks'' contains unit tests for
various C++ helper classes, including concurrent data structures, ``divine'' contains unit tests for a concurrent hashset implementation used in \divine{}, ``cryptopals'' contains solutions of the
cryptopals problem set\footnote{\url{http://cryptopals.com}}, the ``llvm'' category contains programs from the \llvm{} test-suite\footnote{\url{http://llvm.org/svn/llvm-project/test-suite/trunk/SingleSource/Benchmarks/Shootout}}
and finally, the ``svc-pthread'' category includes pthread-based C programs from the SV-COMP benchmark set. In most of the programs, it was assumed that \texttt{malloc} and \texttt{new} never fail,
with the notable exception of part of the ``bricks'' category unit tests. The tests where \texttt{new} failures are allowed are especially suitable for evaluating exception code, in particular where
multiple concurrent threads are running at the time of the possible failure.

\subsection{Comparison to Builtin Exception Support}\label{sec:cmpD3}

In addition to the approach presented in this paper, we have implemented the approach described in~\citep{rockai16:model.checkin} in the context of \divine{}~4. This allowed us to directly measure the
penalty associated with the present approach, which is more thorough and less labour-intensive at the same time. Our expectation was that this would translate to slower verification, since the
off-the-shelf code is more complex than the corresponding hand-tailored version used in~\citep{rockai16:model.checkin}. In line with this expectation, we set the criterion of viability: we would
consider a slowdown of at most 10\,\% to be an acceptable price for the improved verification fidelity, and convenience of implementation. Since other resource consumption (especially memory) of
verification is typically proportional to state space size, we have used the number of states explored as an additional metric. The expected effect on the shape (and, by extension, size) of the state
space should be smaller than the effect on computation time (most of the additional complexity is related to computing a single transition). We believe that an acceptable penalty in this metric would
be about 2\,\% increase.

\hypertarget{tbl:D4D3}{}
\begin{longtable}[]{@{}llllll@{}}
\caption{\label{tbl:D4D3}Comparison of the new exception code with a \divine{}-3-style version. }\tabularnewline
\toprule
category & \#mod & time (D4) & time (D3) & states (D4) & states (D3)\tabularnewline
\midrule
\endfirsthead
\toprule
category & \#mod & time (D4) & time (D3) & states (D4) & states (D3)\tabularnewline
\midrule
\endhead
alg & 9 & 3:52 & 3:51 & 543.3 k & 543.3 k\tabularnewline
pv264 & 13 & 1:34 & 1:32 & 183.0 k & 183.0 k\tabularnewline
iv112 & 11 & 25:58 & 25:57 & 3743 k & 3743 k\tabularnewline
libcxx & 425 & 42:15 & 42:09 & 2182 k & 2182 k\tabularnewline
bricks & 292 & 3:04:25 & 2:56:55 & 6271 k & 6251 k\tabularnewline
divine & 3 & 6:20 & 6:18 & 1040 k & 1040 k\tabularnewline
cryptopals & 3 & 0:01 & 0:01 & 1943 & 1943\tabularnewline
llvm & 12 & 36:36 & 36:27 & 3865 k & 3865 k\tabularnewline
svc-pthread & 17 & 16:47 & 16:41 & 1685 k & 1685 k\tabularnewline
\textbf{total} & 794 & 5:21:44 & 5:13:49 & 20.1 M & 20.0 M\tabularnewline
\bottomrule
\end{longtable}

As can be seen in Table~\ref{tbl:D4D3}, the time penalty on our chosen model set is very acceptable -- just shy of 2.6\,\% -- and the state space size is within 1\,\% of the older
approach~\citep{rockai16:model.checkin}. We believe that this small penalty is well justified by the superior verification properties of the new approach.

\subsection{Comparison to Stub Exceptions}\label{comparison-to-stub-exceptions}

The second alternative approach is to consider any thrown exception an error, regardless of whether it is caught or not. This can be achieved much more easily than real support for exceptions, since
we can simply replace the entire \texttt{libunwind} interface with stubs which raise an error and refuse to continue. This approach only works for models which do not actually throw any exceptions
during their execution. The results of this comparison are shown in Table~\ref{tbl:D4stub} -- the verification time is nearly identical and the state spaces are entirely so. This is in line with
expectations: in those models, catch blocks are present but never executed. Since the proposed approach does not incur any overhead until an exception is actually thrown, we would not expect a
substantial time difference.

\hypertarget{tbl:D4stub}{}
\begin{longtable}[]{@{}lllll@{}}
\caption{\label{tbl:D4stub}Comparison of the new exception code against stubbed exceptions. Compared to Table~\ref{tbl:D4D3}, in this case 133 models failed due to the stubs. State counts are
identical for all models. }\tabularnewline
\toprule
category & \#mod & time (D4) & time (stub) & states (D4)\tabularnewline
\midrule
\endfirsthead
\toprule
category & \#mod & time (D4) & time (stub) & states (D4)\tabularnewline
\midrule
\endhead
alg & 9 & 3:52 & 3:52 & 543.3 k\tabularnewline
pv264 & 13 & 1:34 & 1:34 & 183.0 k\tabularnewline
iv112 & 11 & 25:58 & 26:00 & 3743 k\tabularnewline
libcxx & 392 & 41:56 & 41:54 & 2179 k\tabularnewline
bricks & 192 & 35:30 & 35:21 & 2378 k\tabularnewline
divine & 3 & 6:20 & 6:19 & 1040 k\tabularnewline
cryptopals & 3 & 0:01 & 0:01 & 1943\tabularnewline
llvm & 12 & 36:36 & 36:28 & 3865 k\tabularnewline
svc-pthread & 17 & 16:47 & 16:43 & 1685 k\tabularnewline
\textbf{total} & 661 & 2:52:30 & 2:52:08 & 16.2 M\tabularnewline
\bottomrule
\end{longtable}

\subsection{Comparison to No Exceptions}\label{comparison-to-no-exceptions}

Finally, the last alternative is to disable exception support in the C++ frontend entirely. In \texttt{clang}, this is achieved by compiling the source code with the \texttt{-fno-exceptions} flag. In
this case, the \llvm{} bitcode contains no exception-related artefacts at all, but many programs fail to build. Additionally, a number of programs in the ``bricks'' category contain exception handlers
for memory allocation errors\footnote{In this case, the handler is installed using \texttt{std::set\_terminate}, which is available even when \texttt{-fno-exceptions} is given. The situation would be
  similar if only parts of the program were compiled with \texttt{-fno-execptions}. In particular, the problem is that the standard library, if compiled with \texttt{-fno-exceptions}, cannot throw,
  and must therefore behave differently in those scenarios, affecting the behaviour of the user program.} and therefore exit cleanly upon memory exhaustion. Even though some of those programs can be
compiled with \texttt{-fno-exceptions}, they now contain an error (a null pointer dereference) which is not present when they are compiled the standard way. Those programs were therefore excluded from
the comparison. The summary of this comparison can be found in Table~\ref{tbl:D4nxc} -- the time saved for models where \texttt{-fno-exceptions} is applicable is again quite small, less than 13\,\%.
In this case, the difference is due to the changes in control flow of the resulting \llvm{} bitcode. Since \texttt{call} is not a terminator instruction (unlike \texttt{invoke}), the \emph{local} control
flow in a function is negatively affected by the presence of \texttt{invoke} instructions: more branching is required, and this slows down the evaluator in \divm{}. While it is easy to see if a given
program can be compiled with \texttt{-fno-exceptions}, it is typically much harder to ensure that its behaviour will be unchanged. For this reason, we do not consider the time penalty in verification
of this type of programs a problem.

\hypertarget{tbl:D4nxc}{}
\begin{longtable}[]{@{}lllll@{}}
\caption{\label{tbl:D4nxc}Comparison of the new exception support against a case where \texttt{-fno-exceptions} was used to compile the sources and libraries. In this case, it was only possible to
verify 423 models from the set (i.e.~371 models are missing from the comparison). State counts are identical for all models. }\tabularnewline
\toprule
category & \#mod & time (D4) & time (nxc) & states\tabularnewline
\midrule
\endfirsthead
\toprule
category & \#mod & time (D4) & time (nxc) & states\tabularnewline
\midrule
\endhead
alg & 1 & 0:24 & 0:23 & 34.2 k\tabularnewline
pv264 & 1 & 0:00 & 0:00 & 57\tabularnewline
iv112 & 10 & 23:58 & 22:06 & 3571 k\tabularnewline
libcxx & 393 & 41:57 & 40:44 & 2180 k\tabularnewline
svc-pthread & 17 & 16:47 & 15:42 & 1685 k\tabularnewline
\textbf{total} & 423 & 1:23:33 & 1:19:21 & 7504 k\tabularnewline
\bottomrule
\end{longtable}

\subsection{Re-usability}\label{re-usability}

As outlined in Section~\ref{sec:components}, the two components directly involved in exception support are comparatively small and well isolated. The \llvm{} transformation is fully re-usable with any
\llvm{}-based tool. The unwinder, on the other hand, relies on the capabilities of \divm{}. However, there is no need for hypercalls specific to exception handling and therefore, the implementation work is
essentially transparent to \divm{}. The capabilities of \divm{} required by the unwinder are limited to the following: linked-list stack representation, runtime access to the program bitcode and 2
hypercalls: \texttt{\_\_vm\_control} and \texttt{\_\_vm\_obj\_free}. More details about \divm{} can be found in~\citep{rockai18:divm}.

Finally, adding support for a new type of exceptions is also much simpler in this approach -- no modifications to \divm{} (or any other host tool) are required: only the two components described in this
paper may need to be modified.

\section{Conclusion}\label{sec:conclusion}

In this paper, we have discussed an approach to extending an \llvm{}-based model checker with C++ exception support. We have found that re-using an existing implementation of the runtime support library
is a viable approach to obtain complete, standards-compliant exception support. A precondition of this approach is that the verification tool is flexible enough to make stack unwinding possible. The
\divm{} language, on which the \divine{} model checker is based, has proven to be a good match for this approach, due to its simple and explicit stack representation, along with a suitable set of control
flow primitives.

We also performed a survey of tools based on partial or complete reimplementations of C++ exception support routines and found that in each tool, at least one edge case is not well supported. In
contrast to this finding, with our approach, all those edge cases are covered ``for free'', that is, by the virtue of re-using an existing, complete implementation. Contrary to the prediction made
in~\citep{rockai16:model.checkin}, we have found that with a suitable target language, implementing a new unwinder can be relatively simple. The unwinder implementation described in this paper is only
about 350 lines of C++ code, while it would be impossible to implement without verifier modifications in \divine{}~3. Therefore, we can conclude that with the advent of the \divm{}
specification~\citep{rockai18:divm} and its implementation in \divine{} 4, re-implementing the \texttt{libunwind} API and re-using \texttt{libc++abi} became a viable strategy to provide exception
support.

\bibliography{common}

\end{document}